# Experimental search for dynamic heterogeneities in molecular glass formers

**F Ladieu , C Thibierge and D L'Hôte**
Service de Physique de l'Etat Condensé (DSM/DRECAM) and URA CNRS , CEA Saclay, 91191 Gif-sur-Yvette, France

**Abstract.** We have measured the linear dielectric susceptibility of two molecular glass formers close to $T_g$ in order to estimate the size of the dynamically correlated clusters of molecules which are expected to govern the physics of glass formation. This size has been shown to be related to the dynamic dielectric susceptibility $d\varepsilon(\omega)/dT$ ($\varepsilon$: dielectric susceptibility, $T$: temperature, $\omega$: frequency). To allow for an accurate determination of the $T$ derivative, we scanned the interval $192 < T < 232$ K every 1 K for glycerol and $159 < T < 179$ K every 0.5 K for propylene carbonate. The resolution on $T$ variations was about 1 mK. The result for glycerol is that the number of correlated molecules increases by a factor 3 when $T$ goes from 226 to 195 K. It has been shown that the non-linear susceptibility provides a direct measurement of dynamic correlations. To measure it, we used a standard Lockin technique yielding the third harmonic of the current flowing out of a capacitor. We obtained only an upper limit on the ratio of the third to the first harmonic, due to the non-linear response of standard electronics.

## 1. Introduction

The structural α-relaxation time of supercooled liquids exhibits a fast non-Arrhenius temperature dependence which is among the most striking and yet unexplained feature of glass formers [1-12]. Despite its unsolved origin, this $T$-dependence is expected to be related to cooperative effects [9-13]: The dynamics of the system slows down as the temperature decreases to the glass transition temperature $T_g$ because larger and larger numbers of molecules have to move in a correlated way to allow for the motion of any single molecule. The idea of such cooperatively rearranging domains has been put forward long ago [13-14], and was reinforced by numerical simulations results [14-21]. On the experimental side, correlation lengths have been extracted, using various techniques, which lead to length scales of the order of 5 to 20 particles [9-11, 22-26].

In this paper, we present two experiments devoted to the extraction of the average number of correlated particles in a glass former, $N_{\text{corr}}$, by using two new methods based on linear [27] and non-linear [28] dielectric spectroscopy. These methods allow for a precise determination of the temperature dependence of $N_{\text{corr}}$. We stress the experimental specific problems raised by each of the two methods. The linear dielectric spectroscopy needs an accurate temperature dependence, thus a control of the reproducibility and time stability. The method we used for non-linear dielectric spectroscopy needs low level harmonic distortion of the electronics used.

## 2. Experimental set-up

The experiments were performed in a cryogenerator whose base temperature $T_{min}$ is 10 K. The experimental stage was a closed metallic cell (diameter = 10 cm, height = 3.5 cm) placed in vacuum and connected to the low temperature stage by a thermal impedance ($\approx$ 30 K/W). Since we intend to determine the thermal derivative of the complex dielectric susceptibility $\varepsilon(\omega)$, see equation (1) below, we paid special attention to the measurement of the temperature as well as to the reproducibility and time stability of all our $\varepsilon(\omega)$ measurements (glycerol and propylene carbonate). The temperature $T$ of the experimental cell was regulated by using a PID LakeShore® controller, which gives a stability of $T$ better than 1 mK. The value of $T$ was measured with a resistive thermometer, previously calibrated with a refined method using the triple point of water and the liquefaction temperature of Nitrogen. The thermometer was immersed in the supercooled liquid as close as possible to the samples. The samples were thin layers of glass formers sandwiched between the parallel electrodes of the capacitors. The spacers were small disks (surface 10 mm²) of Kapton or Mylar of thickness $L \approx$ 30 µm (for glycerol) to 360 µm (for propylene carbonate). For the glycerol measurements, electrodes were placed horizontally and were made of highly doped silicon wafers whose surface (20 cm²) was coated with 0.4 µm of gold. For the propylene carbonate measurements, electrodes were placed vertically and were made of brass squares with a polished surface (5.5 cm²). The electrodes were immersed in the glass former, and the filling of the cell was done in an inert atmosphere of Argon. A pressure of 2 bars of Argon was put in the cell at room temperature, thus the pressure above the supercooled liquid remained larger than 1 bar at low temperature in order to avoid the formation of bubbles of adsorbed gases. The voltage $V(\omega)$ was applied to the sample and the resulting current $I(\omega)$ was deduced from the lock-in measurement of the voltage $v(\omega)$ across a 1 k$\Omega$ resistor put in series with the sample. Coaxial shielding was ensured from room temperature down to the experimental cell. The relation between $I(\omega)$ and $v(\omega)$ involves the complex impedances of the cables connecting the sample: The later were carefully measured in order to correct for their contribution. We verified that the stray capacitances were extremely small, if any (at most a few pF). The overall consistency was checked by measuring well known capacitors and resistors on the frequency interval reported here. Finally, the complex capacitance $C(\omega)$ of the sample was deduced from $j\omega C(\omega)=I(\omega)/V_s(\omega)$, where $V_s(\omega)$ is the voltage really applied to the sample (once the whole circuit has been taken into account).

## 3. Linear dielectric spectroscopy experiments

Using this experimental setup, the dielectric response of glycerol (purity 99.6%) was measured at low electric field $E \approx$ 50 kV/m. In the following, we use the notations: $\varepsilon(\omega=0) = \varepsilon'(\omega=0) = \varepsilon(0)$ and $\varepsilon(\omega=\infty) = \varepsilon'(\omega=\infty) = \varepsilon_\infty$ where $\varepsilon'(\omega)$ is the real part of $\varepsilon(\omega)$. Figure 1 shows, for a set of temperatures above $T_g \approx$ 190 K, the real part of $[C(\omega)-C(\infty)]/[C(0)-C(\infty)]$ which is equal to $[\varepsilon'(\omega)-\varepsilon_\infty]/\Delta\varepsilon$ where $\Delta\varepsilon = \varepsilon(0) - \varepsilon_\infty$. Contrary to $\varepsilon(0)$, we do not directly measure the value of $\varepsilon_\infty$, but it can be deduced from fitting our $C(\omega,T)$ data with the Havriliak-Negami parametrization [29]. It turns out that $\varepsilon_\infty/\varepsilon(0)$ is a small number whose $T$-dependence is sufficiently weak to be of no consequence hereafter (see equation (1) below). In our setup, the crystallization of glycerol could start above 210 K in a time scale comparable to that of experiments, yielding distorted curves for both $Re[C(\omega)]$ and $Im[C(\omega)]$. Below 205K this did not occur: the $C(\omega)$ curves (i.e. both $Re[C(\omega)]$ and $Im[C(\omega)]$) were reproducible within 0.1% for days. Starting from 204 K, a few $C(\omega,T)$ curves were recorded at higher temperatures before coming back to 204 K, recording again at 204 K the $C(\omega)$ values and repeating the cycle for different higher temperatures. This allows to discard *a posteriori* the curves where crystallization had started. The curves on figure 1 are those for which the error due to crystallization is less than 1%.

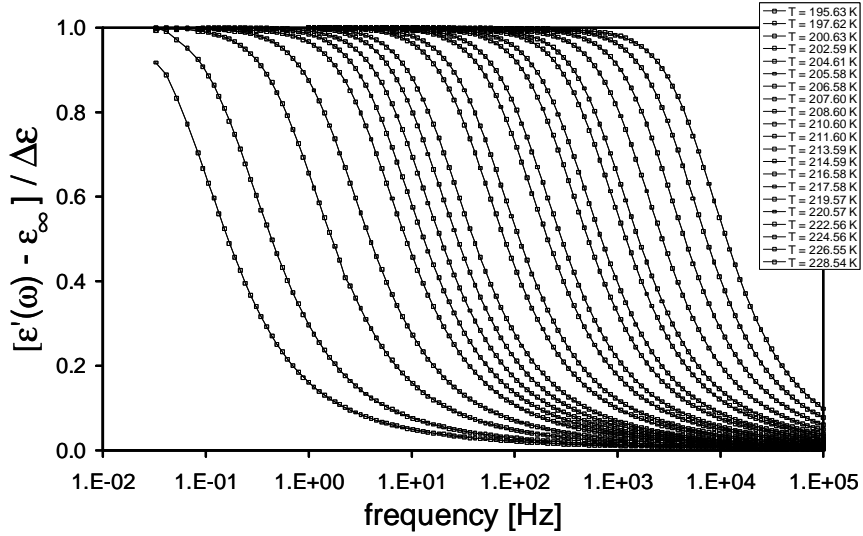

**Figure 1.** The real part of the dielectric susceptibility (normalized by $\Delta\varepsilon = \varepsilon(0) - \varepsilon_\infty$) measured for glycerol in the supercooled state. The temperature goes from 195.63 K (left curve) to 228.54 K (right).

Our set of curves of figure 1 can be used to estimate the number $N_{corr}$ of correlated molecules. Indeed, defining $\chi = Re[\varepsilon(\omega)-\varepsilon_\infty]/\Delta\varepsilon$, it was established recently [27] that

$$N_{corr}(\omega,T) \approx \frac{k_B}{c_P}\left(T\frac{d\chi}{dT}\right)^2 \quad , \tag{1}$$

where $k_B$ is the Boltzmann constant, $c_P$ is related to the specific heat (of a molecular volume). For consistency with reference [27] we have taken for $c_P$ the specific heat jump at $T_g$ (i.e. the difference between the specific heat of the supercooled liquid and that of the crystal). Note that if we had used the specific heat of the supercooled liquid, which is about twice the specific heat jump, the values of $N_{corr}(\omega,T)$ would have been halved. Besides, the right hand side of equation (1) was obtained in reference [27] by considering a four point correlation function whose integral over space gives $N_{corr}(\omega,T)$ up to a numerical prefactor. Here, for consistency with reference [27] we have taken this factor equal to 1. This choice, as well as that concerning $c_p$ (and its possible $T$-dependence that we neglect), will be discussed in a future publication.

To obtain $N_{corr}(\omega,T)$ from our data, we used two methods: (i) we fitted $C(\omega,T)$ with the Havriliak-Negami parametrization [29], and fitted the $T$-dependence of these parameters to finally calculate $d\chi/dT$ ; or (ii) we directly calculated $d\chi/dT$ from finite differences of the curves shown in figure 1. Figure 2 shows that both methods yield similar results. At a given $T$, $N_{corr}(\omega,T)$ reaches its maximum at a frequency $\omega^*$ which is close to the frequency $\omega_\alpha$ at which the imaginary part $\varepsilon''(\omega)$ of the dielectric susceptibility is maximum. For clarity, the full set of $N_{corr}(\omega,T)$ curves is shown only for method (ii), while for method (i) only the maximum value (over $\omega$) of $N_{corr}(\omega,T)$ - called $N_{corr}*(T)$ - is reported for each $T$ . We estimated our experimental uncertainty on the quantity $(T\, d\chi/dT)^2$ (which appears in equation (1)) to vary from 1% for the maximum values of $N_{corr}$ to 5% when $N_{corr} \approx 1$. The overall experimental uncertainty on $N_{corr}$ is larger because it includes the uncertainty on $c_P$ which is rather of the order of 10%. In figure 2, it is worth noticing that method (ii) yields "clean curves" despite the fact tat no fitting procedure of the data is involved. This is consistent with the error bar of at most 1% on $N_{corr}*(T)$. Analyzing figures 2 and 3 in detail reveals that the finite difference method for $T$ steps of 1 K underestimates $N_{corr}$ by typically 7%. The weak difference between the two methods (see figure 3) comes from the finite size of the $T$ steps in method (ii).

Our main result is the increase of $N_{corr}*$ when $T$ decreases towards $T_g$. The analysis of this behavior should yield further insight in the understanding of the glass transition.

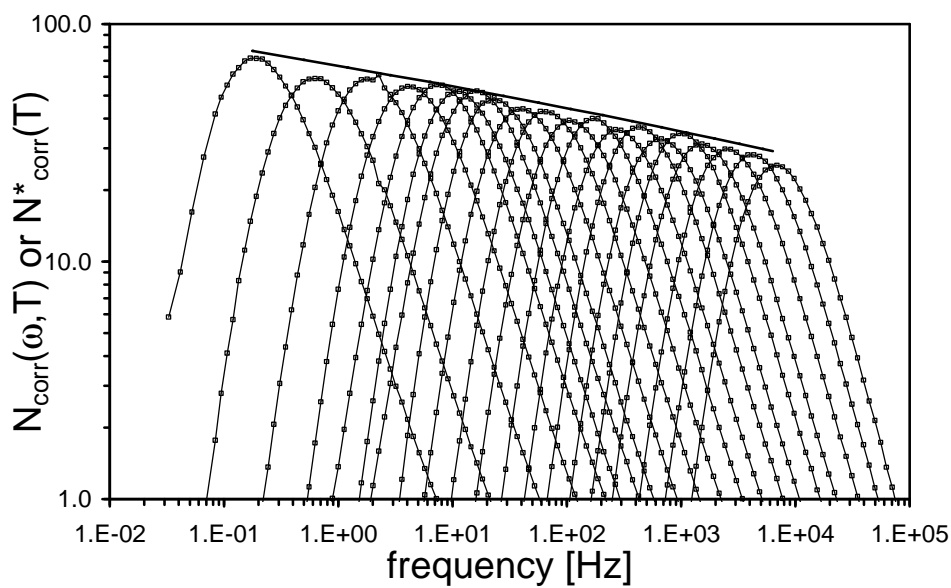

**Figure 2.** Number $N_{corr}(\omega,T)$ of correlated molecules in glycerol obtained from the data shown on figure 1 and using equation (1). The series of parabolic-like curves with squares symbols corresponds to the method (ii) (which does not involve any fitting procedure, see text). By using method (i), the upper solid line is obtained, which corresponds to $N_{corr}*(T)$ defined in the text as the maximum values, over $\omega$, of $N_{corr}(\omega,T)$ for each $T$.

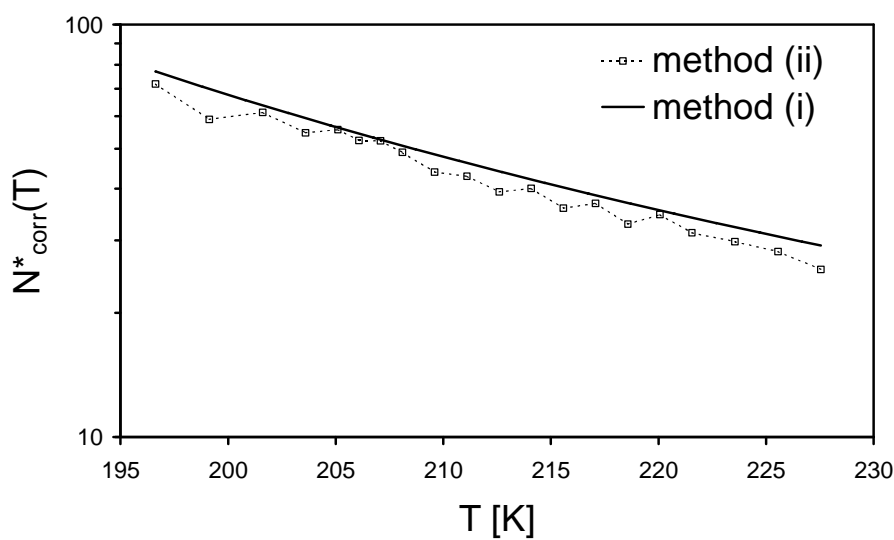

**Figure 3.** Maximum value $N_{corr}*(T)$ of $N_{corr}(\omega,T)$, as a function of $T$, obtained either with method (i) or (ii) (see text for details).

Figure 4 shows the temperature and frequency dependence of the dielectric susceptibility of propylene carbonate. The non-Arrhenius nature of this very fragile glass [1-12] is visible on the figure since the interval between two successive curves increases when the temperature decreases.

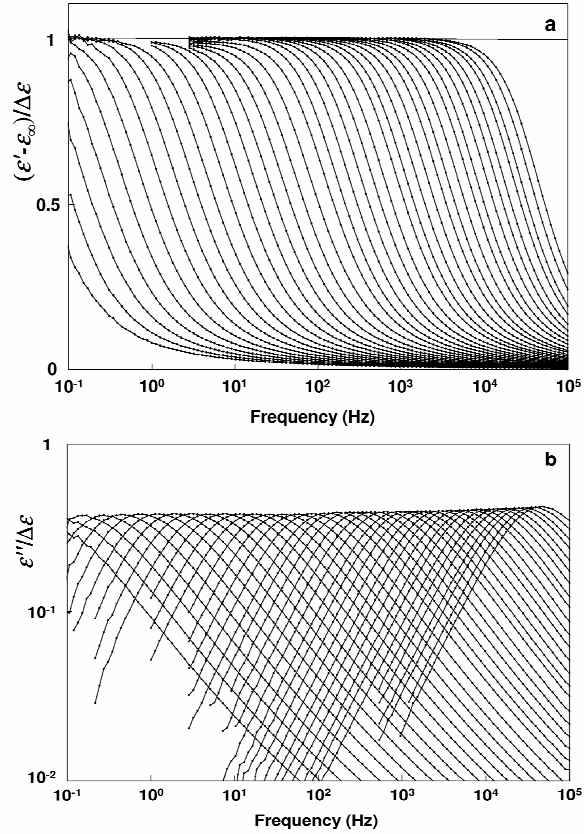

**Figure 4**. **a**: The normalized real part of the measured dielectric susceptibility of propylene carbonate as a function of the frequency for temperatures ranging from 159.34 K to 178.34 K by steps of 0.5 K. **b**: The same as (**a**), but for the normalized imaginary component of the dielectric susceptibility. The normalization consists in dividing $\varepsilon'$ and $\varepsilon''$ by $\Delta\varepsilon = \varepsilon(0) - \varepsilon_\infty$.

### 4. Non-linear dielectric spectroscopy experiments

On quite general theoretical grounds, it has been shown recently [28] that the nonlinear susceptibility $\chi_3(\omega,T)$ is directly related to the cooperative length $l$ by:

$$\chi_3(\omega,T) = \frac{\chi_s^2}{k_B T} l^{2-\eta} H(\omega\tau_\alpha) \qquad , \qquad (2)$$

where $\chi_s$ is the static dielectric susceptibility (not to be confused with the $\chi$ of equation (1)), $\eta$ an exponent related to the spatial structure of the four point correlation function, $H$ a function which should depend only on $\omega\tau_\alpha$ where $\omega$ is the frequency of the perturbing field, and $\tau_\alpha = 1/\omega_\alpha$. $H$ is expected to reach its maximum (of order 1) at $\omega\tau_\alpha \sim 1$, while for $\omega\tau_\alpha \gg 1$ and $\omega\tau_\alpha \ll 1$, $H$ should be much smaller.

Experimentally when an electric field is applied to the sample, the non linear response makes that a third harmonic current $I(3\omega) \sim \chi_3(\omega,T)E^3$ is added to the current $I(\omega)$. By using standard lock-in techniques [30], we looked for $I(3\omega)$ with glycerol samples. For each temperature between 220K and 203K, the frequency $\omega$ was selected to meet the condition $\omega\tau_\alpha \sim 1$, at which the function $H$ should be maximum. We used a standard source (voltage V $\leq 10$ V ) and for the thinnest sample (thickness $L = 30$ µm) the field $E = V/L$ ranged from 10 to 220 kV/m. For a given $L$, we detected a third harmonic current $I(3\omega) \sim V^3$. However, by varying $L$ we found that the expected $I(3\omega) \sim 1/L^3$ law was *not* obeyed. We therefore conclude that the measured $I(3\omega)$ is dominated by a spurious contribution from the electronics. This spurious contribution may come, e.g. from the fact that any voltage source has some harmonic distortion: in addition to the voltage $V(\omega)$, the source delivers small harmonic voltages $\delta V(n\omega)$ with $n = 2,3,...$ The main features relevant for our study are that, typically $\delta V(n\omega) \sim [V(\omega)]^n$ and that $\delta V(3\omega)/V(\omega)$ can reach $10^{-4}$ for the maximum voltages. These considerations explain fairly well that the measured $I(3\omega)$ was dominated by the contribution of the source. As a consequence, we obtained *an upper limit* for the third harmonic current and thus an upper bound for $N_{corr} \approx l^3/a^3$ ($a^3$ being the molecular volume): assuming $H(\omega\tau_\alpha=1)=1$, we deduce an upper bound on $N_{corr}$ typically *ten times larger* than the values reported in figures 2-3. Current work is in progress to reduce the spurious contribution of the electronics.

Finally, we note that Richert and Weinstein [31] have recently performed non-linear measurements on glycerol. Their data are accounted for by a model where both the absorption of electrical energy and the thermal coupling to the bath are governed by the same time scale heterogeneously distributed in the sample. To disentangle this effect from that of equation (2) remains an open question.

**Acknowledgements**
This work was triggered by a collaboration between Saclay (CEA) and Montpellier (CNRS and university) and it is a pleasure to thank our colleagues L Berthier, L Cipeletti, D El Masri, M Pierno, G Biroli, J-P Bouchaud. We also acknowledge stimulating discussions with C Alba-Simionesco, G Tarjus, R Richert and A Reiser. We thank P Pari and P Forget for their help in cyogenics, and R Tourbot for the wonderful realization of the experimental cell.